# Analysis of Performance Parameters in Wireless Networks by using Game Theory for the non Cooperative Slotted Aloha Enhanced by ZigZag Decoding Mechanism

Abdellah ZAALOUL and Abdelkrim HAQIQ
Computer, Networks, Mobility and Modeling laboratory FST, Hassan 1st University, Settat, Morocco
e-NGN research group, Africa and Middle East
{zaaloul, ahaqiq}@g mail.com

Abstract— In wireless communication networks, when the workload increases, sources become more aggressive at the equilibrium in the game setting in comparison with the team problem by using slotted Aloha mechanism. Consequently, more packets are in collision and are lost. To reduce these phenomena and to enhance the performance of the networks, we propose to combine ZigZag decoding approach with non cooperative slotted Aloha mechanism. This approach was introduced in our previous work based on the cooperative slotted Aloha mechanism. The obtained results showed that this approach has significantly improved the cooperative slotted Aloha mechanism and gave best results for the throughput and delay.

In this paper, we analyze the impact of combining non cooperative slotted Aloha and ZigZag Decoding. We model the system by a two dimensional Markov chain that integrates the effect of ZigZag decoding. The states of the Markov chain describe the number of backlogged packets among the users. We use a stochastic game to achieve our objective; we evaluate and compare the performances parameters of the proposed approach with those of a simple slotted Aloha mechanism. All found results show that our approach improves the performance parameters of the system.

## I. INTRODUCTION

Medium access control is a distributed approach in access to a shared wireless channel among competitive nodes. Since wireless networks use a shared transmission medium, collision may occur because of simultaneous transmissions by two or several interfering nodes, hence the necessity to coordinate transmissions. The richest family of multiple access protocols is probably the Aloha family of protocols [1], Carrier Sense Multiple Access (CSMA) [2], and their corresponding variations have been widely studied as efficient methods to coordinate the medium access among competing users. To resolve contention, in these mechanisms, each user either maintains a persistent transmission probability or adjusts a backoff window. For example, using slotted ALOHA mechanism; to reduce contention a user transmits a packet with certain probability during each time slot. But we can see in CSMA mechanism, a user maintains a back off window and waits for a random amount of time bounded by the back off window before a transmission (or retransmission). We had extensively studied the performance of these mechanisms of random access from system perspective, where mobile users are considered as homogeneous devices that always follow the transmission protocol. Wireless nodes usually are not exactly aware of number of nodes in network and each node can obtain some limited information about channel state (e.g. collided packets, busy or idle state of channel) through listening to channel. In such conditions the best thing a node can do, is to optimize its personal goals. Therefore, for modeling such situation, non-cooperative game models are the best choice.

Recently, the users are considered as intelligent and rational individuals who make decisions to maximize their own benefits [3] [4] [5] [6] [7],hence, many studies start to look at the random access problem from the user perspective. In these works, game theory has been applied to modeling and analyzing the random access process with autonomous and heterogeneous mobile users. As a powerful tool to study the interactions among intelligent and rational individuals, game theory has the potential to provide insights and analytical approach to the design of efficient random access mechanisms. In general, the game theoretic approach for analyzing random access contains three steps: game formulation, equilibrium analysis, and mechanism design. The first step is to model the random access process as a game. Typically, the mobile users are considered as players, and their transmission decisions are the actions. Each user is also assigned an utility function, which characterizes the user's satisfaction of a particular outcome (i.e., the utility can be the gain from a successful access minus the transmission cost). Under different assumptions on the users' decision making process (i.e., private objective and available information of other users), the random access game can be formulated as a single shot game with deterministic strategy [4], or a game with probabilistic strategy [3], or a Bayesian game [5]. Once the random access game is formulated, the next step is to analyze the equilibrium of the game. Specifically, the existence of Nash equilibrium and corresponding conditions are analyzed in this step. The equilibrium analysis provides significant insights on whether the system can be operated at a stable state such that users do not change their decisions unilaterally. Finally, based on the equilibrium analysis, an efficient random access mechanism can be designed to guide users to operate at a desired equilibrium state.

This paper studies game theoretic approaches for random access in wireless networks improved by introducing ZigZag Decoding [8]. A time slotted system is considered, where time is divided into slots, and users make their decisions to access a shared channel at the beginning of each time slot. The proposed study of random access in this paper is where users can choose a retransmission probability when making the access decisions. A new game formulation is provided and the corresponding equilibrium is analyzed. This work may provide insights for designing random access mechanisms for next generation wireless systems. Hence, the objective of this work is to determine the impact of combining game theory model and ZigZag Decoding on the system performances in terms of throughput and minimization expected delays of transmitted and backlogged packets in the next generation networks.

To evaluate performances of the proposed access method, we model the system by a two dimensional Markov chain. Let the first state component to be the number of backlogged packets among the users from the 1st to the Mth user, and the second component is the number of backlogged packets of (M+1)th user.

The Markov chain associated with this algorithm is then studied; its stationary distribution is determined, and the improvement of the average throughput and the average delay is then highlighted.

The rest of the paper is organized as follows: We begin by introducing a brief overview of ZigZag approach in section 2. In section 3 we give a brief overview of related work on random access schemes and ZigZag decoding approaches, in section 4 we describe the proposed protocol. In section 5 we construct a Markov Model for non cooperative slotted Aloha combined with ZigZag decoding where each user seek to maximize its own throughput. In section 6 we formulate a cooperative team problem combined with ZigZag decoding where users maximize the total throughput of the system. Section 7 is devoted to numerically study where we compare the properties of the proposed model with cooperative and non cooperative slotted Aloha. Section 8 concludes the paper.

## II. Zig Zag approach Overview

Gollakota.S defines in [8] that ZigZag is a new decoding technique that increases random access methods resilience to collisions. A great advantage of ZigZag is that it requires no changes to the MAC layer and introduces no overheard in the case of no collision. If no collision occurs, ZigZag acts like a typical random access method. Another important aspect is that ZigZag achieves the same performance as if the colliding packets were a priori scheduled in separate time slots [9]. In ZigZag method, the receiver can decode two consecutive signals of two colliding packets and successfully receive both packets despite collision. In other words, if the same two packets collide twice, the receiver can receive both of those packets. Thus, the maximum achievable throughput of a wireless network can be significantly improved by using ZigZag decoding method. According to [8] there are some basic characteristics for ZigZag decoding method:

1- A ZigZag method can operate with unmodified network structure.
2- ZigZag method decreases the loss rate average.
3- Averaging over all sender-receiver pairs, including those that do not suffer from hidden terminals, the authors find that ZigZag improves the average throughput.

## III. Related Work

In this part of our work, we look some Prior works that has already analyzed this simple approach in many networks systems. Traditionally, a popular solution for wireless network is represented by a random multiple access. The most popular protocol employed and still employed is the slotted Aloha protocol [1, 10]. When multi-users send packets over common channel, random access schemes let to this population of users to share dynamically and opportunistically this channel. In practice, the level of coordination among the users wishing to access the channel is low or impossible (in many scenarios), this may be due to several reasons: for instance, to a lack of global information, to a too large user population size, or to the sporadic and unpredictable nature of users' access activity [11]. So, packets sent at the same time by several users fall in collision. Several works have studied and sought to resolve this collision phenomenon. The authors in [8] with ZigZag decoding show how to recover multiple collided packets in a 802.11 system when there are enough transmissions involving those packets. The main idea of Zig Zag approach is based on interference cancellation, the successive interference cancellation (SIC) technique was employed as a protocol in [12]. In this respect, SIC techniques has turned out to represent a major advance, allowing collisions be favorably exploited instead of being regarded as simply a waste.

Authors in [13] propose a scheme named CRDSA (Contention Resolution Diversity Slotted Aloha) exploiting SIC in the case of satellite access networks to remarkably improve the performance of the diversity slotted Aloha techniques (DSA) [14], this scheme consist of transmitting each packet twice in a medium access control (MAC) frame. The authors in [15] suggest ZigZag decoding, to extract the packets involved in the collisions. They present an algebraic representation of collisions and describe a general approach to recovering collisions using

Analog Network Code ANC. They studied the effects of using ANC on the performance of MAC layers. The use of ZigZag without additional digital network coding has recently been considered by [16] to improve congestion control and maximize aggregate utility of the users.

In another related paper [17], the authors provide an abstraction of the multiple-access channel when ZigZag decoding is used at the receiver. They use this abstract model to analyze the delay and throughput performance of the system. Using various scenarios they conclude that the mean delivery time of the system with ZigZag decoding is strictly smaller than for a system with a centralized scheduler. Otherwise the stability region of the ZigZag decoding system is strictly greater.

In our previous work [23], we studied cooperative slotted system combined with ZigZag decoding. The study show that this approach, improve significantly the team problem performance.

## IV. PROPOSED PROTOCOL

The nature of the wireless network is intrinsically different from the wired network because of the sharing of the medium among several transmitters. Such a restriction generally has been managed through forms of scheduling algorithms to coordinate access to the medium, usually in a distributed manner. The conventional approach to the Medium Access Control (MAC) problem is contention-based protocols in which multiple transmitters simultaneously attempt to access the wireless medium and operate under some rules that provide enough opportunities for the others to transmit.

The Slotted aloha protocol is probably one of the most popular in the multiple access protocols family. It has long been used as random distributed medium access for radio channels. Indeed, it is so simple that its implementation is straightforward, and many local area networks of today implement some variants of Slotted aloha. In these protocols, packets are sent simultaneously by more than one user then they collide. Packets that are involved in a collision are backlogged and retransmitted later.

Original slotted-Aloha Protocol (Fig.1.), is base on the following [1]:

- Time is divided into “slots” of one packet duration.
- When a node has a packet to send, it waits until the start of the next slot to send it.
- If no other nodes attempt transmission during that slot, the transmission is successful.
- Otherwise “collision” occurs, and packets involved in a collision are lost.
- Collided packets are retransmitted after a random delay.
- If a new packet arrives during a slot, it will be transmit in the next slot.
- If a transmission has a collision, node becomes backlogged.
- There are three immediate feedback states: Idle (0), Success (1), Collision (C)

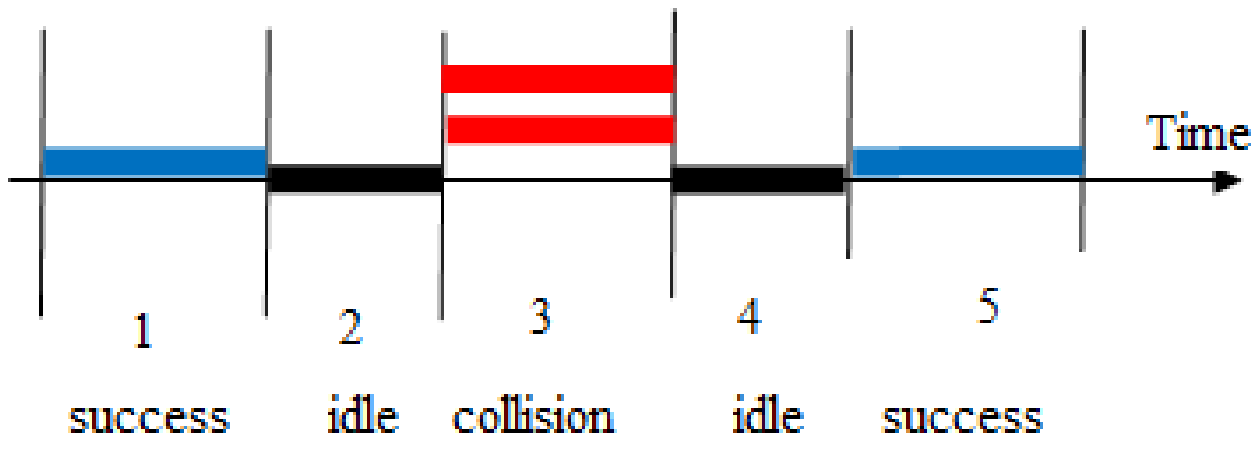

Figure 1: A timeline showing the various kinds of frames for slotted Aloha ‘success’, ‘idle’, ‘collision’.

Collision problem is state traditionally; when two or more packets are transmitted simultaneously a collision occurs. At the MAC layers, many solutions have been tested to eliminate collision hidden and exposed terminals [18].

A proposal is to alleviate the interference impact by learning the interference MAP, and taking scheduling decisions according to this MAP. At higher layers, network coding could also boost the system throughput, as demonstrated in [19].

ZigZag decoding is a new proposed approach for collision resolution [20]. In this approach if the same two packets collide twice, the receiver can receive both of those packets.

In the studied model we propose to combine the slotted Aloha medium access protocol with ZigZag decoding technique.

First the following assumptions are done:

- The frame size either one or two slots.
- At the beginning of a frame, all M users independently transmit (for the first time) or retransmit (in the case of a backlogged user) a packet.

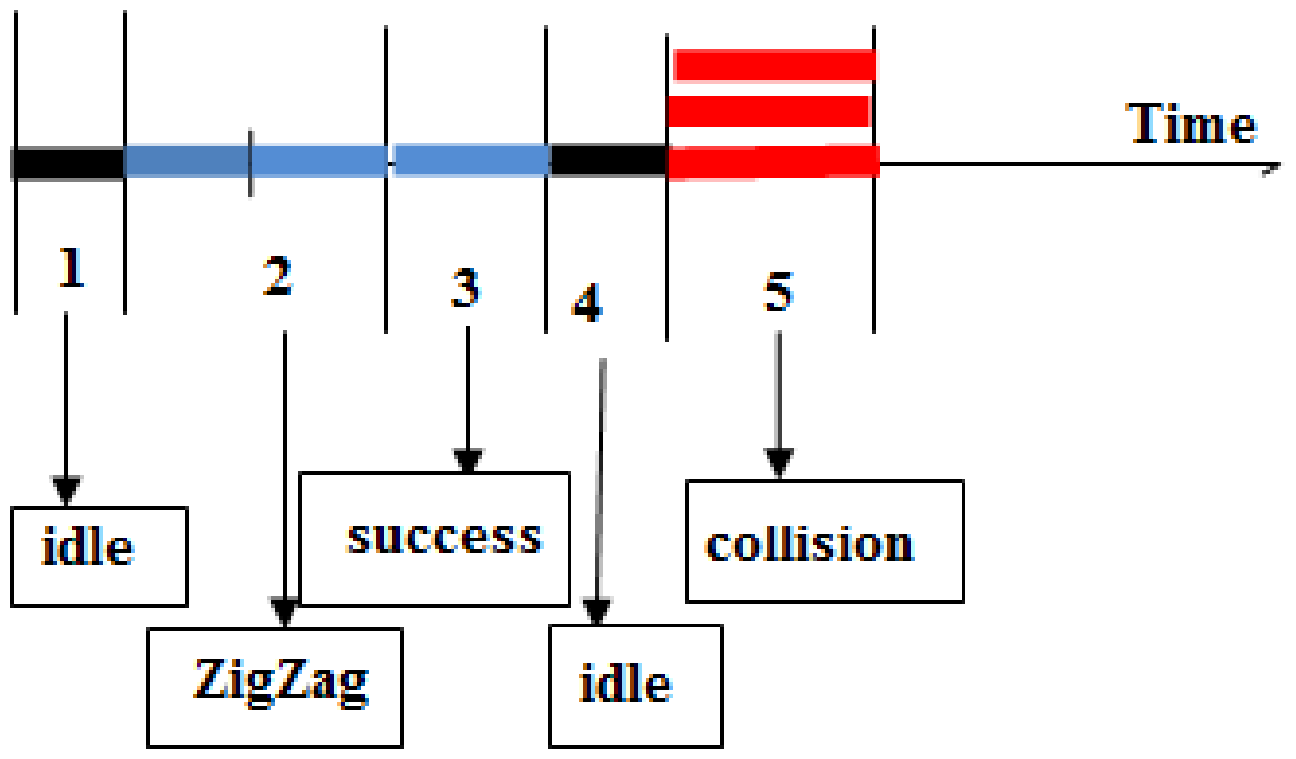

Figure 2: A timeline showing the various kinds of frames for proposed method: ‘success’, ‘idle’, ‘collision’ ‘ZigZag’.

In the proposed protocol, exactly one of the following four events happens (Fig.2.):

1. Idle: nobody transmits any packet,
2. Success: exactly 1 user transmits a packet,
3. ZigZag: exactly 2 users transmit a packet,
4. Collision: when 3 or more users attempt transmission.

Then the receiver gives one of the 4 following feedback messages at the end of the first slot of the frame:

$$Feedback = \begin{cases} 0\ if\ \ idle\ \ (i,e,no\ packets\ attempted\ transmission) \\ 1\ if\ \ success\ (i,e,\ exactly\ 1\ packet\ attempted\ transmission \\ ZigZag\ if\ \ exactly\ 2\ packets\ attempted\ transmission \\ C\ if\ \ 3\ or\ more\ packets\ attempted\ transmission \end{cases}$$

With this scenario the frame has size 1 slot if the feedback is "0", "1", or "C" and if the feedback is "ZigZag" then the frame has a size of 2 slots. Therefore, using ZigZag decoding with Slotted Aloha in our analysis, we can redefine the term "collision" as follow: "A collision occurs on a slot when 3 or more users attempt transmission in a given time slot".

## V. A NON COOPERATIVE GAME

### *A. problem formulation for a non cooperative game*

In many cases slotted Aloha system is usually a decentralized entity, so the cooperative model is not efficient any time, we will develop a model for decentralized non cooperative game which is more powerful and appropriate to slot Aloha. Balance concept replaces the concept of optimality in the team problem.

We consider a cellular system where M nodes transmit over a common channel to a base station.

We denote N the number of backlogged nodes (or equivalently, of backlogged packets) at the beginning of a slot.

For our model, we define and we adopt the following notations:

The arrival flow of packets to source follows a Bernoulli process with parameter $p_a$ (i.e. at each time slot, there is a probability $p_a$ of a new arrival at a source, and all arrivals are independent).

$\overline{Qr}$ : the vector of retransmission probability for all users (whose each entry is $qr$ >0 ),

$\overline{Qa}$: the vector of transmission probability for all users (whose each entry is $p_a$).

Let $Q_a(i,N)$ be the probability that unbacklogged nodes transmit packets in a given slot.

$$Q_a(i,\mathrm{N}) = \binom{M-N}{i}(1-p_a)^{1-M-N}(p_a)^i \qquad (1)$$

And Let $Q_r(i,\mathrm{N})$ be the probability that i out of backlogged nodes retransmit packets in a given slot.

$$Q_r(i,\mathrm{N}) = \binom{N}{i}(1-q_r)^{N-i}(\mathrm{q}_r)^i \qquad (2)$$

We have defined by $\overline{Qr}$ a vector of retransmission probabilities for all users (whose $j^{th}$ entry is $q_r^j$, define $(\overrightarrow{[Q_r]}^j, q_r^j)$ to a retransmission policy where user j retransmit at slot with probability $q_r^j$ for all i≠j and where user i retransmit with probability $q_r^i$. In a non cooperative game, each user $i$ is interested to maximize its own throughput $\mathrm{THp}_i$ .Then the problem that we are interested is to seek a symmetric equilibrium policy $\overrightarrow{q_r^*} = (q_r,...,q_r)$ such that for any user i and any retransmission $q_r^i$ for that user, $\mathrm{THp}_i(\overrightarrow{q_r^*}) \geq \mathrm{THp}_i(\overrightarrow{[Q_r]}^j, q_r^i)$

Next the problem that we seek is to show how to obtain an equilibrium policy. Due to symmetry, to see if $\overrightarrow{q_r^*}$ equilibrium verify the last equation, it suffices to check $\mathrm{THp}_i(\overrightarrow{q_r^*}) \geq \mathrm{THp}_i(\overrightarrow{[Q_r]}^j, q_r^i)$ for a single player. For this, we shall assume that there are $\mathrm{M+1}$ users all together, and that the first $M$ users retransmit with a given probability $\overrightarrow{q_r^*}^{(M+1)}(q^M,...,q^M)$ and user $\mathrm{M+1}$ retransmit with probability $\overrightarrow{q_r}^{(M+1)}$. We can define the set

$$\mathrm{Q}^{M+1}(\overrightarrow{q_r^M}) = \arg\max_{\overrightarrow{q_r}^{(M+1)}} \mathrm{THp}_{M+1}(\overrightarrow{[q_r^M]}^{M+1}, q_r^{(M+1)}) \qquad (3)$$

Where $\overrightarrow{q_r^M}$ denote the policy all users transmit with probability $q_r^M$ , and where the maximization is taken with respect to $q_r^{(M+1)}$ . Therefore, $\overrightarrow{q_r^*}$ is a symmetric equilibrium $\overrightarrow{q_r^*} \in Q^{(M+1)}(\overrightarrow{q_r^*})$.

To compute and compare the performance metrics with studied model in [21] we use again a two dimensional Markov chain improved by ZigZag decoding. Let the first state component to be the number of backlogged packets among the users $1,...,\mathrm{M}$ , and the second component is the number of backlogged packets of user M+1 (either 1 or 0).

$$P_{(N,a)(N+i,b)}(\mathrm{q}_r^M, q_r^{M+1}) = \begin{cases} \left.\begin{array}{ll} Q_a(i,N) & a=b=1 \\ Q_a(i,N)(1-P_a) & a=b=0 \\ Q_a(i,N)P_a & a=0,b=1 \end{array}\right\} 3 \leq i \leq M-N \\ \left.\begin{array}{ll} Q_a(1,N)[1-(Q_r(0,N)+Q_r(1,N))(1-\mathrm{q}_r^{M+1}) & a=b)1 \\ Q_a(1,N)[1-Q_r(0,N)-Q_r(1,N)](1-\mathrm{P}_a) & a=b=0 \\ Q_r(1,N)*Q_a(1,N)*\mathrm{P}_a & a=0\quad b=1 \end{array}\right\} i=1 \\ \left.\begin{array}{ll} Q_a(2,N)[1-(Q_r(0,N))(1-q_r^{M+1})] & a=b=1 \\ Q_a(2,N)[1-Q_r(0,N)](1-P_a) & a=b=0 \\ Q_r(2,N)*\mathrm{P}_a & a=0,b=1 \end{array}\right\} i=2 \\ \left.\begin{array}{ll} (1-q_r^{M+1})ZigZag + q_r^{M+1}(1-Q_r(0,N)-Q_r(1,N))Q_a(0,N) & \mathrm{a=b=1} \\ (1-P_a)ZigZag + P_a[Q_a(0,N).Q_r(0,N)+Q_a(1,N).Q_r(0,N)] & a=b=0 \\ Q_a(0,N).P_a[1-Q_r(0,N)-Q_r(1,N)] & \mathrm{a=0,\ b=1} \\ q_r^{M+1}[Q_a(0,N).Q_r(0,N)+Q_a(1,N).Q_r(0,N)] & a=1,b=0 \end{array}\right\} i=0 \\ \left.\begin{array}{l} (1-q_r^{M+1}[Q_r(1,N).Q_a(0,N)+Q_r(1,N).Q_a(1,N)]\ \mathrm{a=b=1} \\ [Q_r(1,N).Q_a(0,N)+Q_r(1,N).Q_a(1,N)](1-\mathrm{P}_a)\ a=b=0 \end{array}\right\} i=-1 \\ \left.\begin{array}{ll} (1-q_r^{M+1})[Q_r(2,N).Q_a(0,N)\ \mathrm{a=b=1} \\ Q_r(2,N).Q_a(0,N)(1-P_a) & a=b=9 \end{array}\right\} i=-2 \\ 0 \quad otherwise \end{cases} \qquad (4)$$

where:

$ZigZag = [Q_r(0,N).Q_a(1,N) + Q_r(0,N).Q_a(2,N)] + Q_a(0,N)[1 - Q_r(2,N) - Q_r(1,N)]$

Once again, the steady state distribution is solution of the following problem:

$$\begin{cases} \pi((\overrightarrow{[q_r^{M+1}]}^{(M+1)}, q_r^{M+1})) = \pi((\overrightarrow{[q_r^{M+1}]}^{M+1}, q_r^{M+1}))P(p_a, q_r) \\ \pi_{N,i}[\overrightarrow{q_r^{M+1}]}^{(M+1)}, q_r^{M+1}) \geq 0 \; N = 0 \cdots M; i = 0,1 \\ \sum_{N+0}^{M} \sum_{i=O}^{1} \pi_{N,1}(\overrightarrow{[q_r^{M+1}]}^{(M+1)}, q_r^{M+1}) = 1 \end{cases} \quad (5)$$

*B. Performance metrics for game problem*

After we can calculate the number of backlogged packets of user M+1, it is given by:

$$S_{(M+1)}(\overrightarrow{[q_r^{M+1}]}^{(M+1)}, q_r^{M+1}) = \sum_{N=0}^{M} N.\pi_{N,1}(\overrightarrow{[q_r^{M+1}]}^{(M+1)}, q_r^{M+1}) \quad (6)$$

The average throughput of user M+1 is given by:

$$THp_{M+1}(\overrightarrow{[q_r^{M+1}]}^{(M+1)}, q_r^{M+1}) = P_a.\sum_{N=0}^{M} N.\pi_{N,0}(\overrightarrow{[q_r^{M+1}]}^{(M+1)}, q_r^{M+1}) \quad (7)$$

Hence the expected delay of transmitted packets of user M+1 is given by:

$$D_{(M+1)}(\overrightarrow{[q_r^{M+1}]}^{(M+1)}, q_r^{M+1}) = 1 + \frac{S_{(M+1)}(\overrightarrow{[q_r^{M+1}]}^{(M+1)}, q_r^{M+1})}{THp_{M+1}(\overrightarrow{[q_r^{M+1}]}^{(M+1)}, q_r^{M+1})} \quad (8)$$

## VI. PROBLEM FORMULATION FOR A COOPERATIVE TEAM

In this subsection, we describe a cooperative slotted-Aloha combined with a ZigZag decoding technique and we construct a Markov Model based on [21], from which performance parameters are measured.

Under the same notations that we defined before, we consider M nodes that transmit over a common channel to a base station.

It is obvious that N is a Markov Chain for which the state space is $E = \{0,1,\cdots N\}$

The transition probabilities of the Markov chain are given by:

$$P_{N,N+i} = \begin{cases} Q_a(i,N) \quad 3 \leq i \leq M-N \\ Q_a(1,N)[1 - Q_r(0,N) - Q_r(1,N)] \quad i = 1 \\ Q_a(2,N)[1 - Q_r(0,N)] \quad i = 2 \\ Q_a(0,N)[1 - (Q_r(1,N) + Q_r(2,N)] + \cdots \\ [Q_r(1,N) + Q_r(0,N)].Q_a(1,N) + Q_a(0,N).Q_r(2,N) \; i = 0 \\ Q_a(0,N).Q_r(1,N) \quad i = -1 \\ Q_a(0,N).P_{ZigZag} \quad i = -2 \end{cases} \quad (9)$$

With :

$$P_{ZigZag} = \binom{2}{N}(1 - q_r)^{N-2}(q_r)^2$$

Since the state space is finite and all the states communicate between them the Markovian Chain is ergodic.

Let $\pi(p_a, q_r)$ be the corresponding vector of steady state probabilities where its $N^{th}$ entry $\pi_N(p_a, q_r)$ denotes the probability of $N$ backlogged nodes.

This steady state distribution is solution of the following problem:

$$\begin{cases} \pi(p_a, q_r) = \pi(p_a, q_r)P(p_a, q_r) \\ \pi_N(p_a, q_r) \geq 0, N = 0, \ldots M \\ \sum_{N=0}^{M} \pi_N(p_a, q_r) = 1 \end{cases} \quad (10)$$

By computing recursively the steady state probabilities, we can obtain a solution to this problem, by calculating the performance metrics as in [24].

*A. Maximization of the Global Throughput for team problem*

The throughput of the system is defined as the sample average of the number of packets that are successfully transmitted; it is given almost surely by the constant:

$$\begin{aligned} Th(p_a, q_r) = & \sum_{N=0}^{M} \pi_N(p_a, q_r)[P_{N,N-1} + P_{N,N-2} + Q_a(0,N)Q_r(1,N) + \\ & + Q_a(0,\mathrm{N})\mathrm{P}_{ZigZag} + \pi_0(\mathrm{p}_a, q_\partial)Q_a(1,0)] \\ = & P_a \sum_{N=0}^{M} \pi_N(p_a, q_r)(M - N) \end{aligned} \quad (11)$$

Therefore, we are interested to find an optimal solution of the following problem:

$$\max_{q_r} Th(p_a, q_r) \; s.t \begin{cases} \pi(p_a, q_r) = \pi(p_a, q_r)P(p_a, q_r) \\ \pi_N(p_a, q_r) \geq 0, N = 0, \ldots M \\ \sum_{N=0}^{M} \pi_N(p_a, q_r) = 1 \end{cases} \quad (12)$$

We can also calculate the average number of backlogged packets by:

$$S_B(P_a, q_r) = \sum_{N=0}^{M} \pi_N(p_a, q_r).N \quad (13)$$

Using the formula $\sum_{N=0}^{M} \pi_N(p_a, q_r) = 1$ the throughput can be written as follow:

$$Th(p_a, q_r) = p_a(M - S_B). \quad (14)$$

*B. Minimization of the Delay for the team problem*

We can define the delay as the average time, in slots, that a packet takes from its source to the receiver. By little's formula [22], the delay is given by:

$$D(p_a,q_r)=\frac{Th(p_a,q_r)+S_B(p_a,q_r)}{Th(p_a,q_r)} =1+\frac{S_B(p_a,q_r)}{Th(p_a,q_r)} \quad (15)$$

The analysis of the equations (11) and (15) shows that maximizing the throughput is equivalent to minimizing the average delay of transmitted packets.

### C. Performance measures for backlogged packets for the team problem

An interesting alternative for measuring the performance of the system is to analyze the ability to serve packets awaiting retransmission. It has a great interest especially for real-time applications.

Let $T(p_a,q_r)$ is the average throughput of new packets arrived (crowned with success), so the average throughput for backlogged packets is given by:

$$\overline{T}(p_a,q_r)=Th(p_a,q_r)-T(p_a,q_r)$$

Where $T(p_a,q_r)$ is calculated by:

$$T(p_a,q_r)=\sum_{N=0}^{M}\pi_N(p_a,q_r)Q_a(1,\mathrm{N})+Q_a(2,\mathrm{N})Q_r(0,N) \quad (16)$$

Thereafter we can calculate the expected delay $\overline{D}(p_a,q_r)$ for packets backlogged by applying little's formula [22]. Is given by:

$$\overline{D}(p_a,q_r)=1+\frac{S_B(p_a,q_r)}{\overline{T}(p_a,q_r)} \quad (17)$$

## VII. NUMERICAL RESULTS

We present in this section the numerical results that allow evaluating the performance metrics of system using the proposed method for both cooperative and non cooperative slotted aloha. On the other hand, we analyze and compare these metrics with those of the slotted Aloha taken as reference.

Numerical results show that the impact of ZigZag decoding combined with both the cooperative team and the non-cooperative game problem, on throughput and delay, significantly improve those of slotted Aloha.

In addition, we obtain the retransmission probabilities which solve the team and the game problem with the ZigZag Decoding.

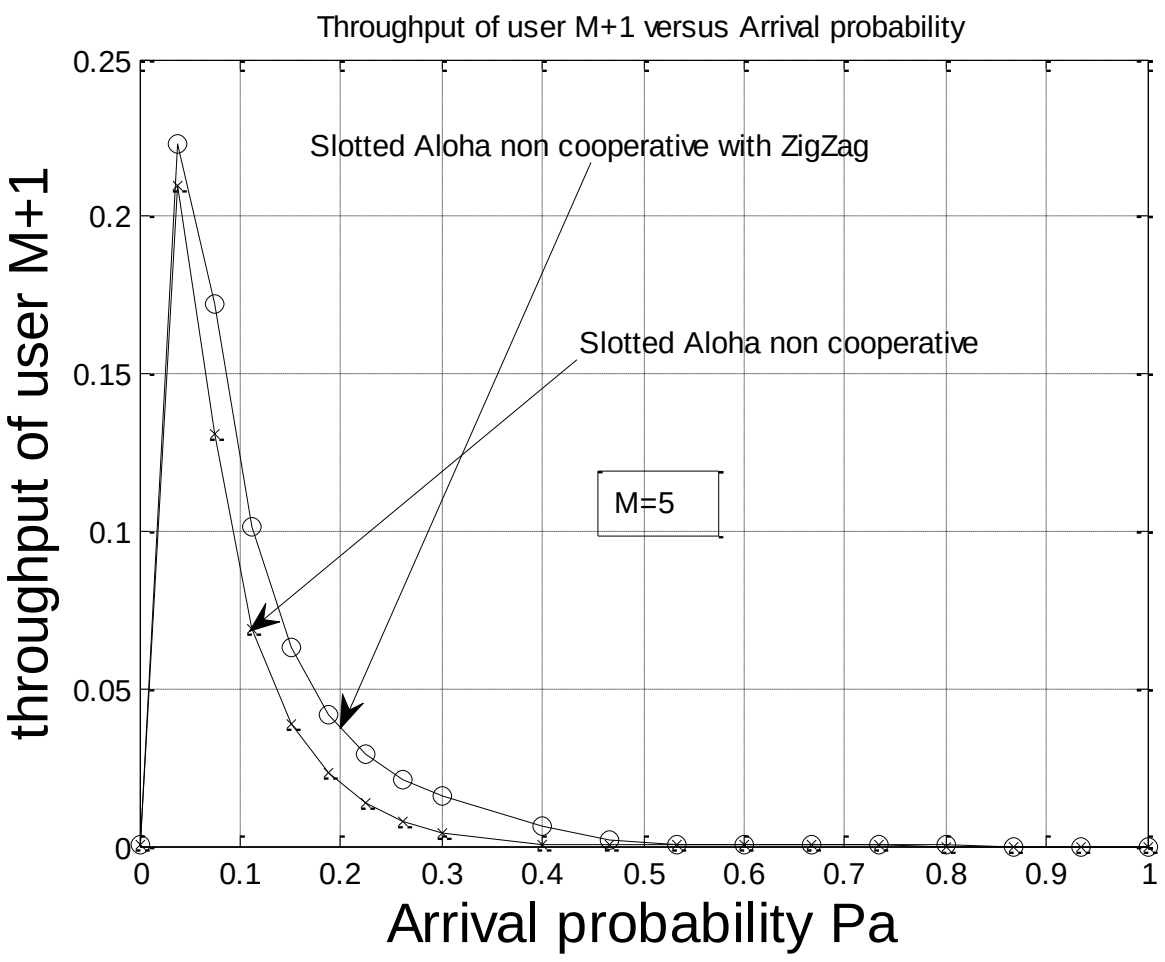

Figure 3: Throughput of user M+1 versus Arrival probability for M=5 in the Game problem.

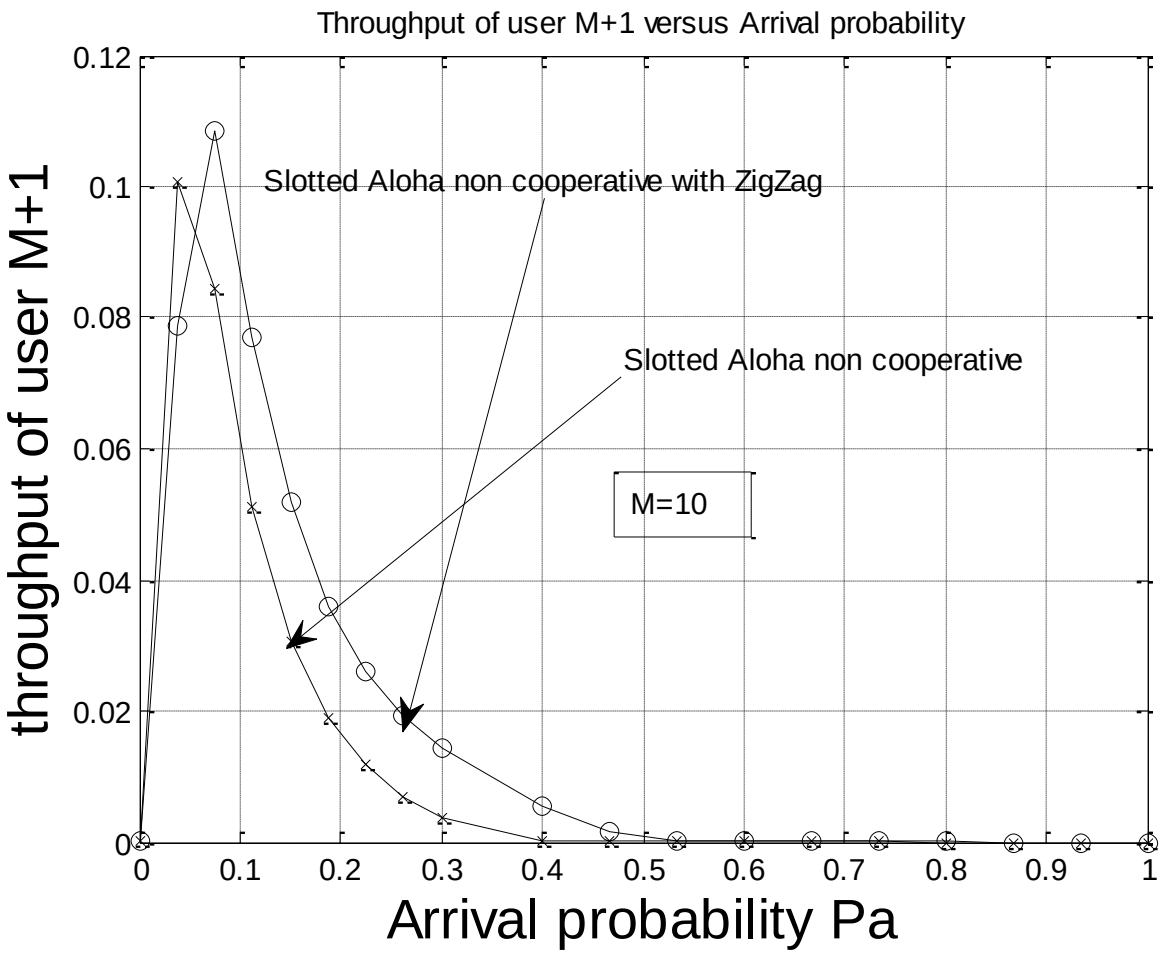

Figure 4: Throughput of user M+1 versus Arrival probability for M=10 in the Game problem.

In Figure 3 and 4 we evaluate the throughput of user M+1 in the game problem. For M=5 and M=10, i.e 5 and 11 mobiles altogether. The equilibrium throughput is a concave function of arrival probability Pa, non cooperative slotted Aloha combined with ZigZag Decoding provides better throughput compared to the implemented non cooperative Slotted Aloha. This comparison is more realistic because the ZigZag Decoding takes into consideration the interferences problem and signal quality needed to decode correctly the captured signal.

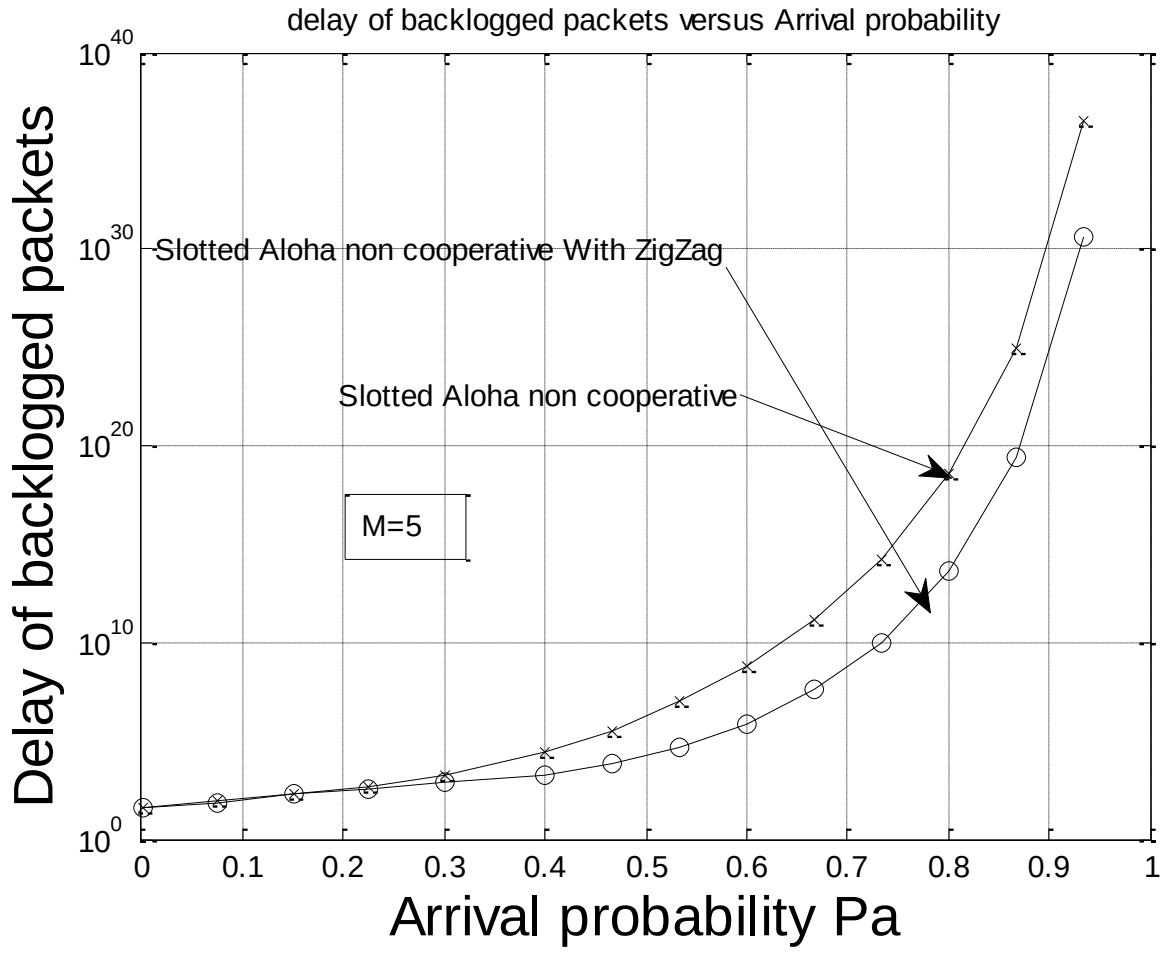

Figure 5: Delay of backlogged packets versus arrival probability for M=5.

This Figure shows that the delay of backlogged packets for non cooperative slotted Aloha with ZigZag is better than delay with slotted non cooperative, except we observe that when Pa< 0.25 both versions have nearly the same performance which is a linear function of Pa. This is due to the fact that there are few newcomers. But when Pa> 0.25 the collision phenomenon has become very important. The effect of ZigZag decoding on reducing collision significantly influenced the delay of backlogged packets.

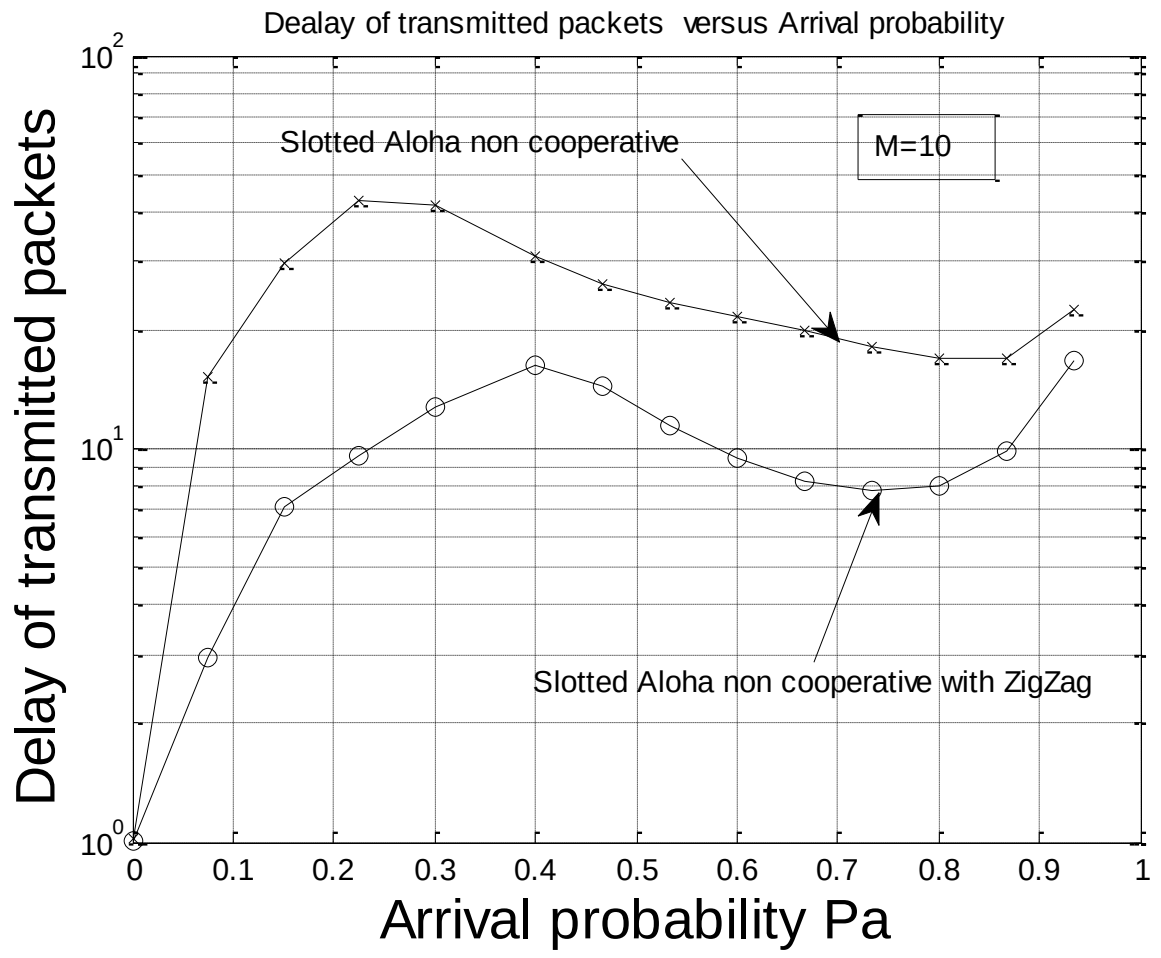

Figure 6: Delay of transmitted packets versus arrival probability for M=10 in the Game problem.

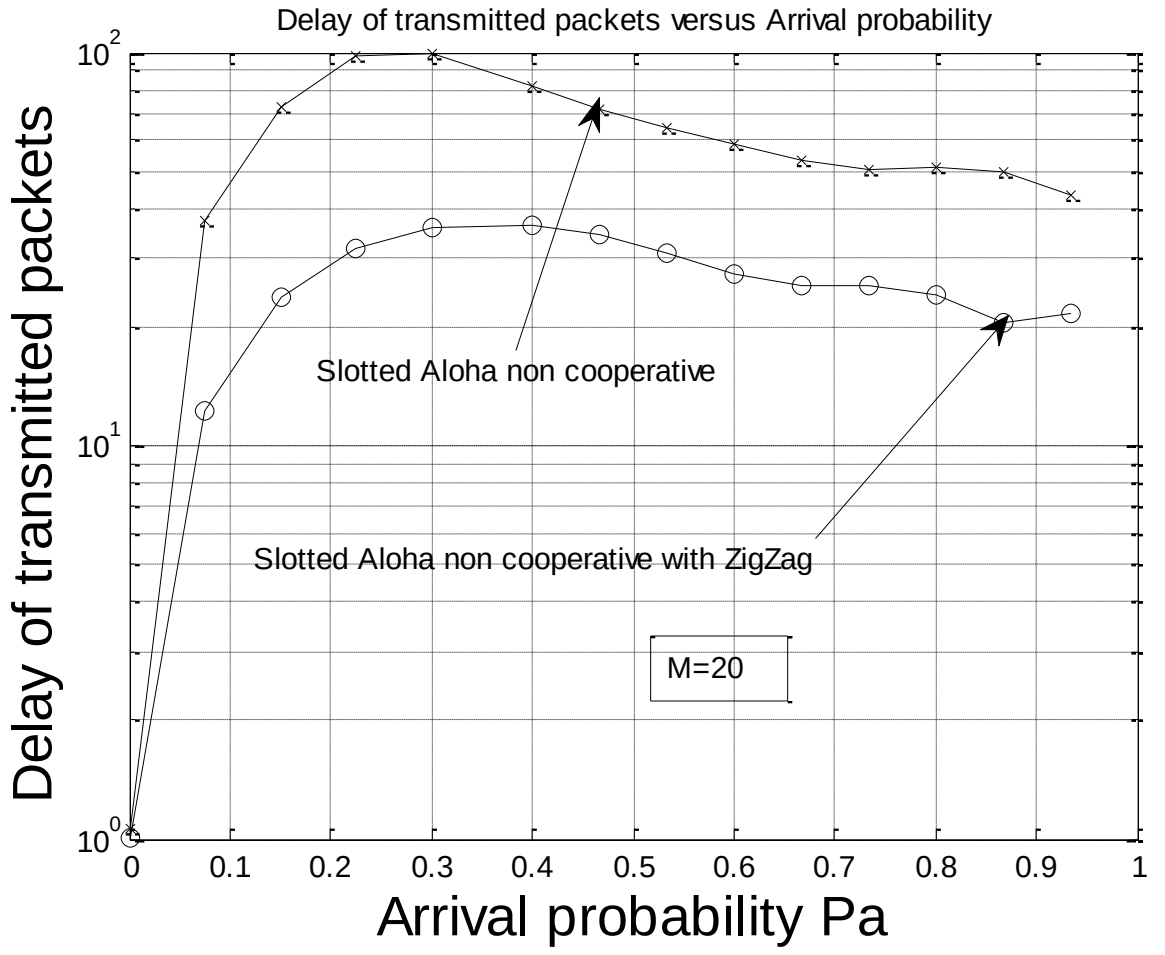

Figure 7: Delay of transmitted packets versus arrival probability for M=20 in the Game problem.

Figure 6 and 7 show the impact of ZigZag decoding on reducing the delay of transmitted packets compared to slotted non cooperative Aloha. ZigZag decoding improves delay, this improvement is important compared to the system without ZigZag decoding. However, this improvement result is not as much as that of the M-user random access model. The reason is because the throughput of system can be improved by ZigZag decoding.

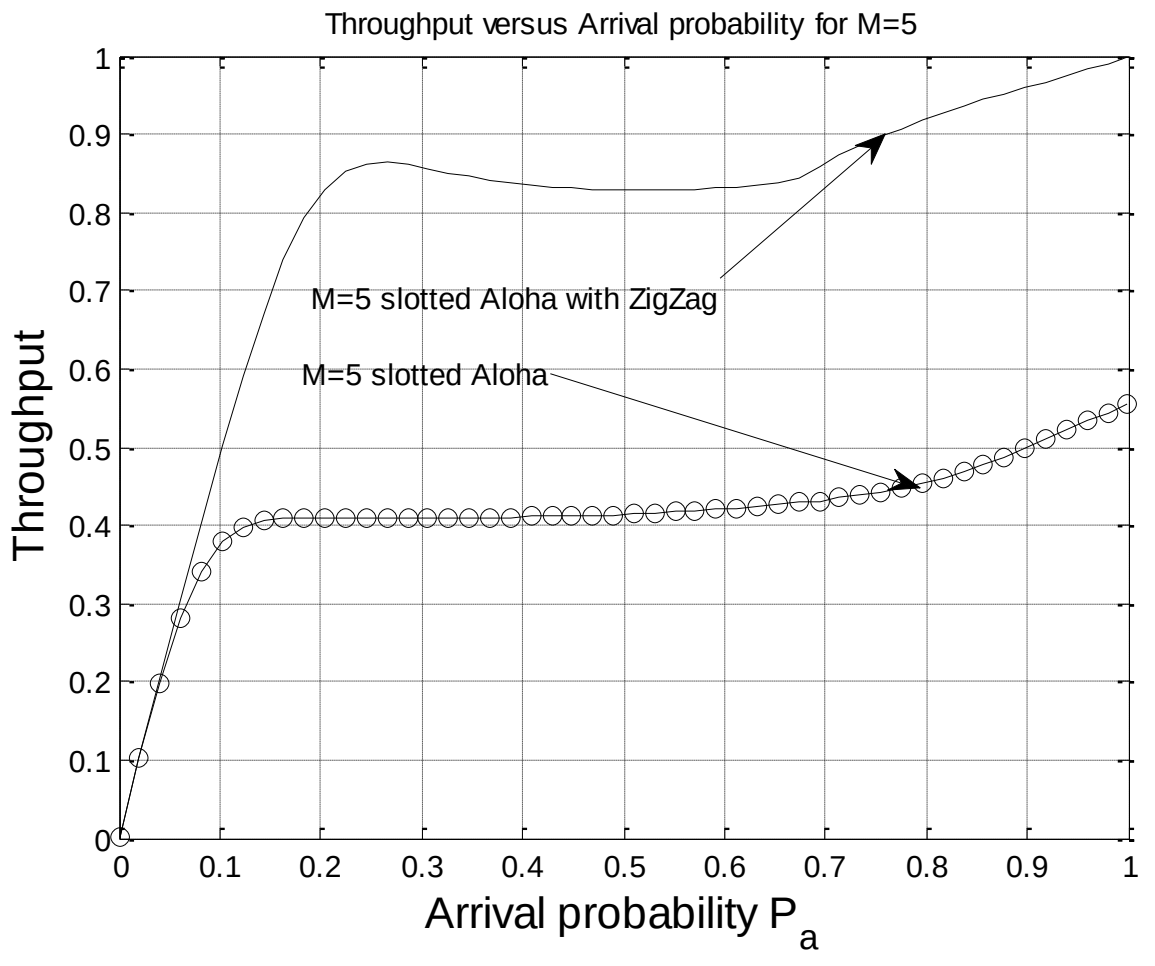

Figure 8: throughput vs arrival probability for M=5 for team problem.

After solving equation (11) for M = 5. We observe that the average throughput, figure 8, has been significantly improved when using the slotted Aloha model with ZigZag decoding, either in low and high traffic; specially when the transmission probability becomes large ( $p_a \succ 0.1$).

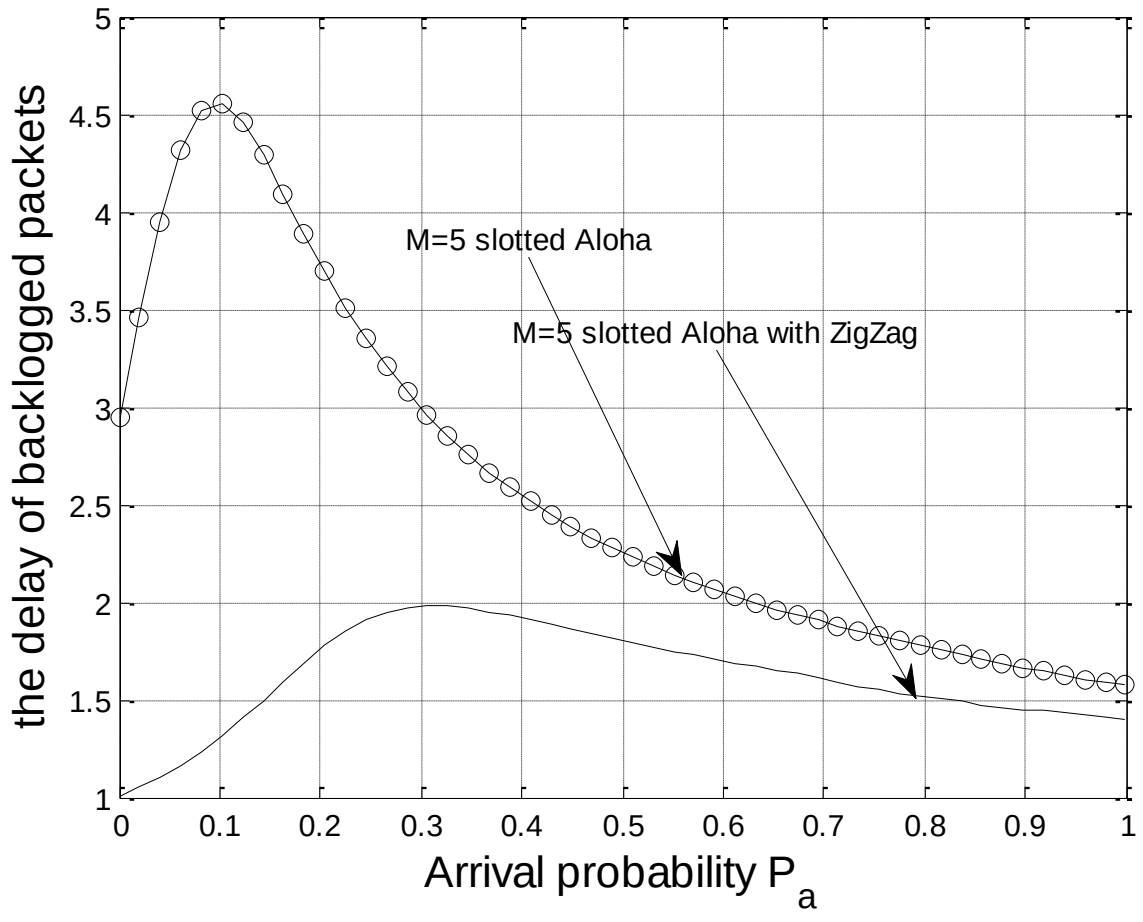

Figure 9: the delay of backlogged packets vs arrival probability for M=5 for team problem

The expected delay of backlogged packets is shown in figure 9. In all cases the proposed method significantly improves the delay of backlogged packets with reference to Slotted Aloha. This improvement is very important when the transmission probability isn't close to 1, and this is true either for heavy or low load.

However, when the load is heavy, we note that the improvement is very clear; the backlogged packets delay is almost reduced by1/3, while the improvement is almost 1/2 at a low load.

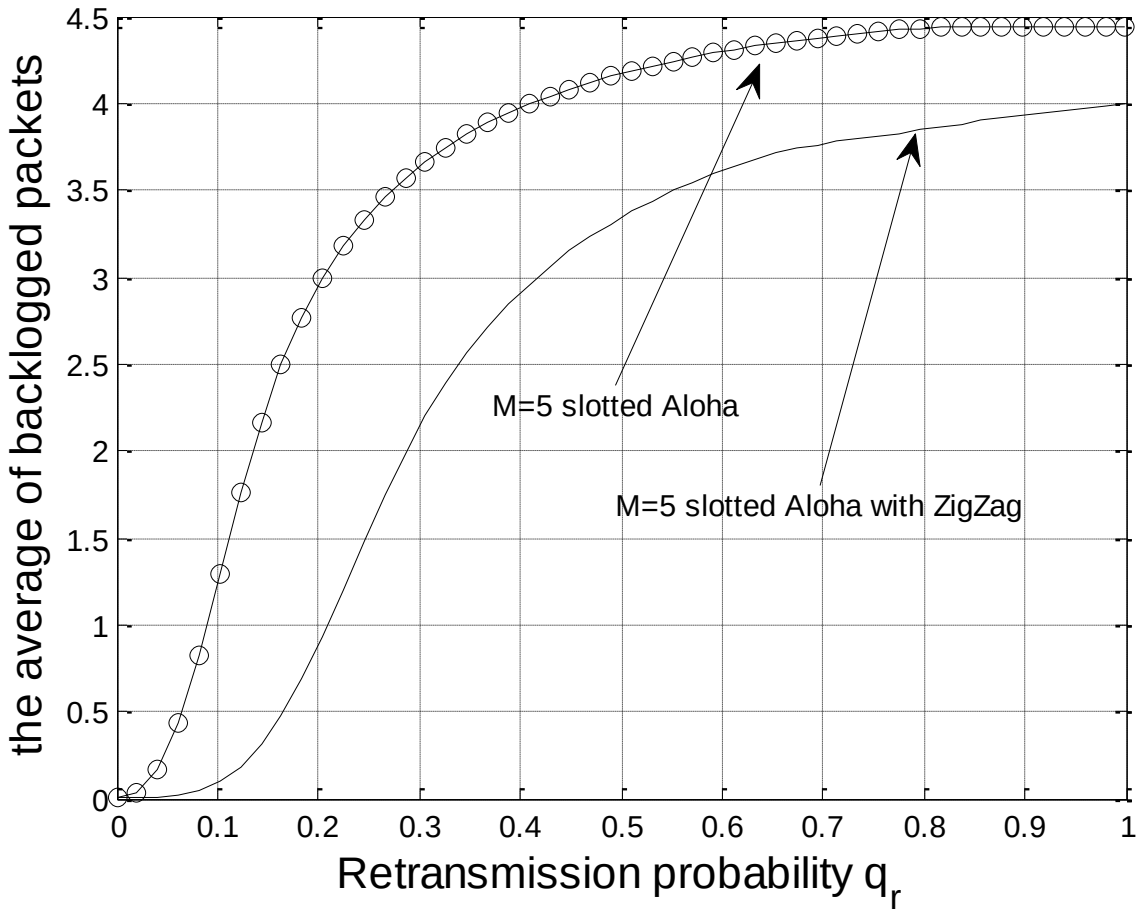

Figure 10: the average of backlogged packets vs retransmission probability for M=5 for team problem.

Figure 10 plot the average of backlogged packets versus the retransmission probability. It show that the new method reduce the number of this type of packets in the system. This improvement is more important when the system load is low.

## VIII. CONCLUSION

In this paper, we have presented a non cooperative slotted Aloha system. We studied the impact of ZigZag decoding on the behavior of non cooperative game so as to improve the minimization of expected delays of transmitted and backlogged packets, and to maximize the system throughput.

The system performances have been evaluated when the arrival probability is varying.

In a first time, we compared the results with a simple non cooperative slotted aloha. The comparison showed that the proposed model improves system throughput and minimize the expected delays of transmitted and backlogged packets.

In second time, the performances are compared with a team problem combined with ZigZag decoding. The results showed some superiorities of the cooperative slotted Aloha combined with ZigZag decoding over the proposed method.

Finally we confirm that ZigZag decoding approach provides better performance when combined with Slotted Aloha. This is more realistic because this algorithm takes into consideration the interferences problem.

AUTHORS PROFILE

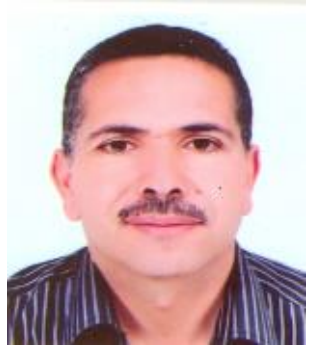

**Abdellah Zaaloul** is currently pursuing his PhD.Degree in the Department of Mathematics and Computer at Faculty of Sciences and Techniques (FSTS), Settat, Morocco. Hi received his M.Sc. degree in Mathematical and Applications engineering from the Hassan 1st University,Faculty of Sciences and Techniques (FSTS), Settat, Morocco, in 2011. And he has been working as professor of mathematics in high school since 1992, Settat, Morocco. Currently. He is member of e-ngn research group His current research interests include performance evaluation and control of telecommunication networks, stochastic control, networking games, reliability and performance assessment of computer and communication systems.

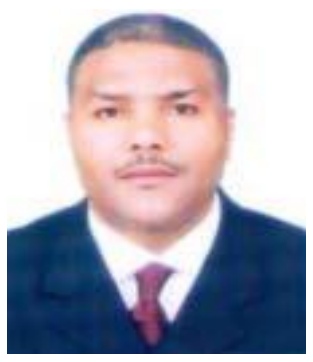

**Dr. Abdelkrim Haqiq** has a High Study Degree (DES) and a PhD (Doctorat d'Etat), both in Applied Mathematics, option modeling and performance evaluation of computer communication networks, from the University of Mohamed V, Rabat, Morocco. Since September 1995, he has been working as a Professor at the Faculty of Sciences and Techniques, Settat, Morocco. He is the Director of IR2M laboratory. He is also a General Secretary of e-Next Generation Networks (e-NGN) Research Group, Moroccan section. Four of his PhD students have presented their thesis between April 2013 and April 2014.
Dr. Abdelkrim HAQIQ is actually Co-Director of a NATO multi-year project entitled "Cyber Security Assurance Using Cloud-Based Security Measurement System" in collaboration with Duke University, USA, Arizona State University, USA and Canterbury University, Christchurch, New Zealand. He is also Co-Director of a Moroccan Tunisian research project entitled "Toword T-Learning based on the Smart TV: A Case Study of the Arab Maghreb" in collaboration with the University of Sfax, Tunisia.
Dr. Abdelkrim HAQIQ's interests lie in the areas of modeling and performance evaluation of communication networks, cloud computing and security. He is the author and co-author of 70 papers (international journals and conferences/workshops). He was a publication co-chair of the fifth international conference on Next Generation Networks and Services, held in Casablanca, May, 28 - 30, 2014. He was also an International Steering Committee Chair and TPC Chair of the international conference on Engineering Education and Research 2013, iCEER2013, held in Marrakesh, July, 1st –5th, 2013, and a TPC co-chair of the fourth international conference on Next Generation Networks and Services, held in Portugal, December, 2 - 4, 2012. Dr. Abdelkrim HAQIQ was the Chair of the second international conference on Next Generation Networks and Services, held in Marrakech, July, 8- 10, 2010. He is also a TPC member and a reviewer for many international conferences. He was also a Guest Editor of a special issue on Next Generation Networks and Services of the International Journal of Mobile Computing and Multimedia Communications (IJMCMC), **July-September 2012, Vol. 4, No. 3, and** a special issue of the Journal of Mobile Multimedia (JMM), Vol. 9, No.3&4, 2014.